\documentclass[a4paper,aip,reprint]{revtex4-1}
\usepackage{graphicx}
\usepackage{amsmath}
\usepackage{fancyhdr}
\usepackage{cancel}
\usepackage{color}
\setcitestyle{round}
\begin{document}
\title{{\it Mini-}grand canonical ensemble: chemical potential in the solvation shell}
\author{Purushottam D. Dixit}
\thanks{Correspondence should be addressed to dixitpd@gmail.com}
\affiliation{Department of Systems Biology, Columbia University}
\author{Artee Bansal}
\affiliation{Department of Chemical and Biomolecular Engineering, Rice University}
\author{Walter G. Chapman}
\affiliation{Department of Chemical and Biomolecular Engineering, Rice University}
\author{Dilip Asthagiri}
\affiliation{Department of Chemical and Biomolecular Engineering, Rice University}

\begin{abstract}
Quantifying the statistics of occupancy of solvent molecules in the vicinity of solutes is central to our understanding of solvation phenomena.  Number fluctuations in small `solvation shells' around solutes cannot be described within the {\it macroscopic} grand canonical framework using a single chemical potential that represents the solvent `bath'. In this communication, we hypothesize that  molecular-sized observation volumes such as solvation shells are best described by  coupling the solvation shell with a mixture of particle baths  each with its own chemical potential. We confirm our hypotheses by studying the enhanced fluctuations in the occupancy statistics of hard sphere solvent particles around a distinguished hard sphere solute particle.  Connections with established theories of solvation are also discussed.
\end{abstract}
\maketitle

A quantitative description of thermodynamics at the nano-scale is of crucial importance in many biological as well as nano-technological systems~\citep{dill2010molecular}. While traditional statistical mecahnical ensembles can describe macroscopic systems, they are inadequate in describing `small systems'. For example, the probability distribution $p(\bar r)$ of degrees of freedom (dof)  $\bar r$ of a small system exchanging energy with a surrounding bath is  {\it not solely} determined by its Hamiltonian $H_{\rm sys}(r)$ and a unique temperature that describes system-bath interactions~\citep{dixit2011elastic,dixit2012role,jarzynski2017stochastic},
\begin{eqnarray}
p(\bar r) \not \propto \exp \left ( -\beta H_{\rm sys}(\bar r) \right ).
\end{eqnarray}

To  describe the distribution of dof accurately, the system Hamiltonian must be augmented by a {\it temperature dependent} potential of mean force $\phi(\bar r; \beta)$ that also depends on the molecular details of the interactions between the system and the bath~\citep{dixit2011elastic,dixit2012role,jarzynski2017stochastic},
\begin{eqnarray}
p(\bar r) \propto \exp \left ( -\beta H_{\rm sys}(\bar r) - \beta \phi(\bar r;\beta) \right ).\label{eq:phi}
\end{eqnarray}
We note that Eq.~\ref{eq:phi} is formally correct but the functional form of $\phi$ in Eq.~\ref{eq:phi} and  its explicit dependence on the bath temperature is seldom known {\it a priori}. 

The molecular field becomes irrelevant for macroscopically large systems (with short range interactions)~\citep{jarzynski2017stochastic}; we expect $\phi(\bar r;\beta) \rightarrow 0$. The failure of the canonical ensemble in describing small systems can be understood by noting that the magnitude of system-bath interactions, $H_{\rm sys-bath}$, is comparable to the magnitude of system-system interactions $H_{\rm sys}$. As a result, small systems cannot weakly couple with a realistic bath~\citep{dixit2013maximum,dixit2015detecting}. 

The inability of small systems to couple weakly to their surroundings is likely to hold true for all statistical mechanical ensembles. In other words, if the system-bath exchanges are comparable to the corresponding property of the system,  statistical mechanics based on average extensive quantities is expected to fail. For example,  the grand canonical framework with a unique chemical potential is likely to be inadequate if the number fluctuations are comparable to the average number of particles in a system.

Recently, we hypothesized that the notion of unique intensive bath parameters  can be relaxed when studying small systems~\citep{dixit2013maximum,dixit2015detecting}.  We showed, using all-atom molecular dynamics (MD) simulations, that the equilibrium properties and dynamics of a small system exchanging energy with its surrounding can be accurately described by a super-statistical generalization of the canonical ensemble wherein the small system is coupled to multiple heat baths each with a different temperature~\citep{dixit2013maximum,dixit2015detecting}.

In this work, we focus our attention on the grand canonical ensemble at the microscopic scale. We study the statistics of number fluctuations in solvation shells of solute molecules. Understanding the number statistics at small length scales is of particular interest in biochemistry;  typically small molecules bind to biological macromolecules such as proteins and nucleic acid polymers (RNA and DNA) in small `binding sites' whose chemical composition can fluctuate. Indeed, the thermodynamics of preference of small molecules over their competitors in such binding sites  directly depends on the statistics of  `ligands'  in the binding site~\citep{dixit2009ion,asthagiri2010ion,dixit2011role,dixit2011thermodynamics}. 

\begin{figure}
	\includegraphics[scale=0.3]{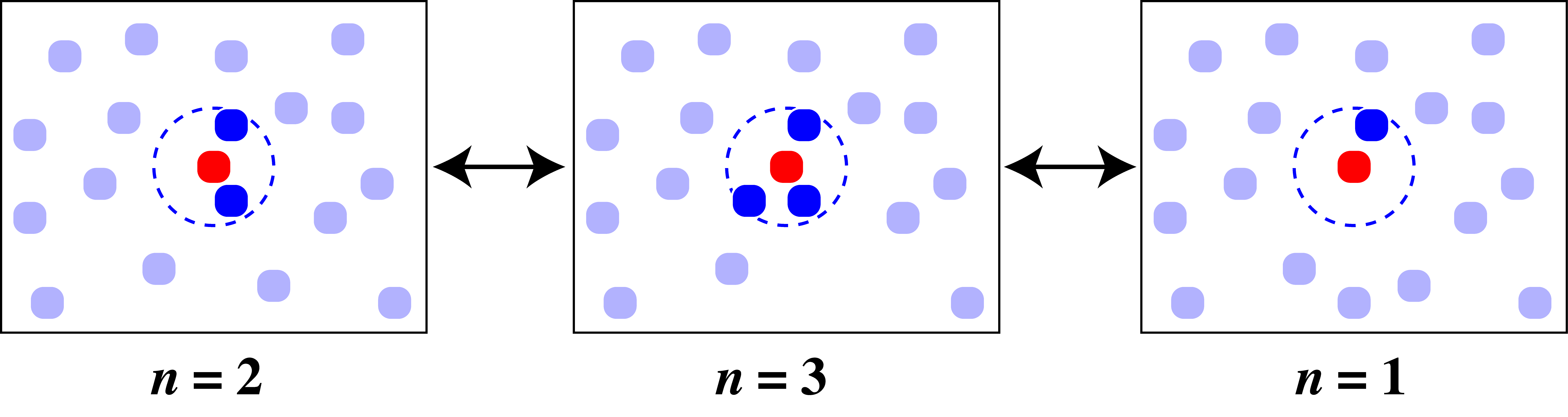}
	\caption{Illustration of the number fluctuations in a `solvation shell' (dashed circle) of a solute (red circle) as the system samples configurations from the microcanonical ensemble.\label{fg:fig1}}
\end{figure}

We work with a hard sphere system to avoid confounding effects due to energetic interactions. For concreteness, we consider a  bath of $N\gg 1$ hard sphere particles with one solute particle fixed at the origin. Let the radius of each hard sphere particle, including the solute, be $r_p$.  Within the bath,  imagine a `solvation shell'  of radius $R$ around the solute. The number $n$ of solvent particles inside the shell fluctuates as the bath samples configurations according to the microcanonical ensemble (see Fig.~\ref{fg:fig1}). The probability $p(n|R)$ of observing $n$ solvent particles in the solvation shell is a centrally important quantity in the study of hydration phenomena, such as ion solvation and the hydrophobic effect~\citep{asthagiri2008role,merchant2009thermodynamically,asthagiri2010ion,bansal_structure_2016,bansal_thermodynamics_2017}.

We note that if $R \gg r_p$, the grand canonical ensemble predicts that the probability of observing $n$ particles in the solvation shell is 
\begin{eqnarray}
p_\mu(n|R) \sim e^{- F(n) +  \mu n}\label{eq:gc0}
\end{eqnarray}
where $F(n)$ is the free energy of assembling $n$ solvent particles around the solute in a shell of radius $R$ {\it in the absence of the rest of the solvent}.   Without loss of generality we have assumed that $\beta = 1$. Here,  $\mu$ is the chemical potential that dictates system-bath coupling. Note that $F(n)$ only depends on the configurations of the system and does not depend on the nature of exchange of particles between the system and the bath and on the chemical potential of the bath. A key feature of the grand canonical description is that a single bath parameter $\mu$ describes all moments of the number distribution $p_\mu(n|R)$ as derivatives of the grand canonical partition function~\citep{dill2010molecular}. 

\begin{figure}
	\includegraphics[scale=0.7]{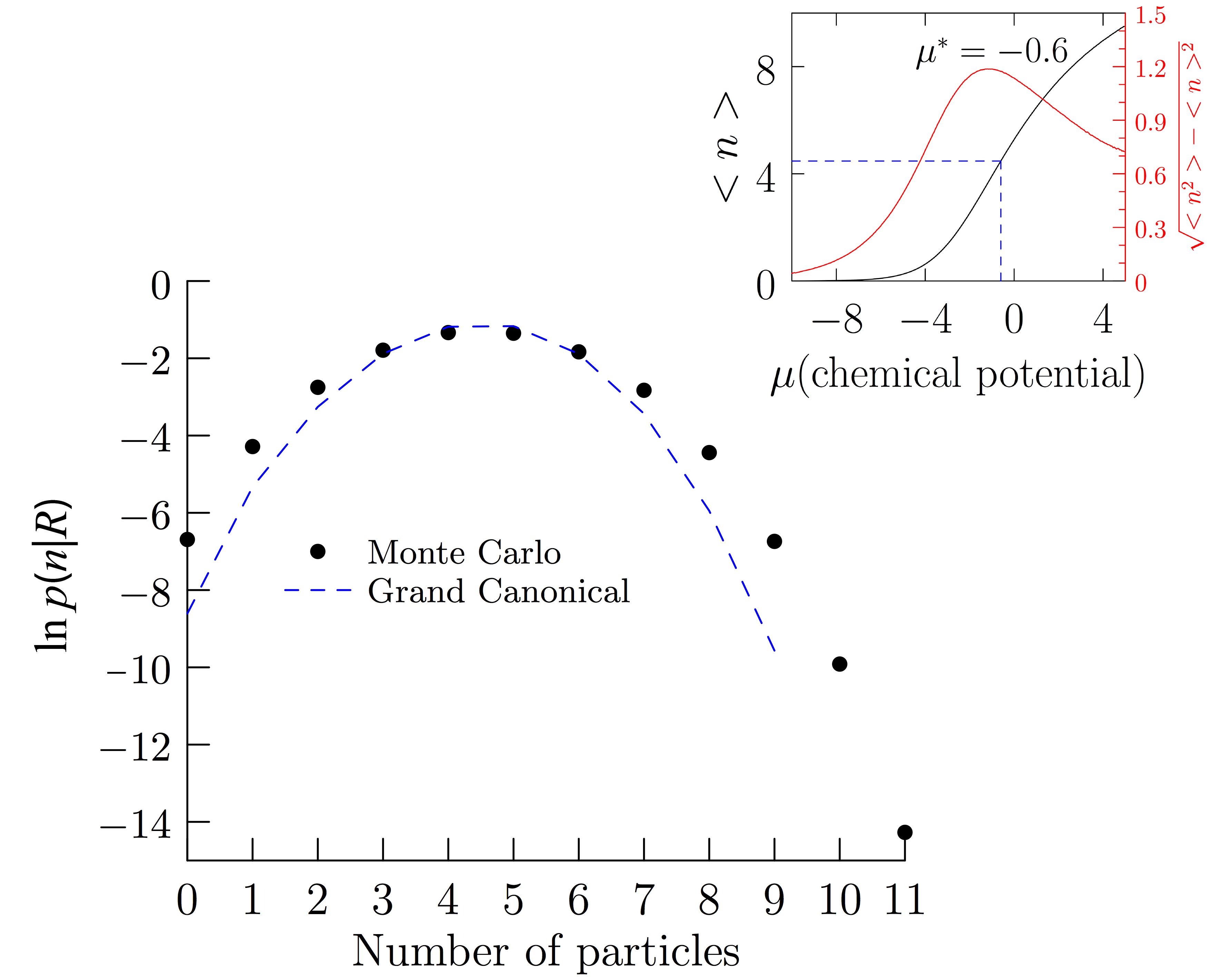}
	\caption{Distribution $p(n|R)$ of the number $n$ of solvent particles inside the solvation shell of radius $R=2.2\times r_{\rm p}$ of a  spherical solute particle with radius $r_{\rm p}$ in a solution of three dimensional hard sphere solvent particles of radius $r$ (black circles). The reduced  density of the solvent is $\rho \sigma^3 = 0.9$. Dashed blue line shows the distribution $p_{\mu = \mu^{*}} (n|R)$ of a grand canonical ensemble simulation where the chemical potential $\mu^{*} \approx -0.6$ is adjusted to reproduce the average occupancy $\langle n \rangle \approx 4.5$. The inset shows the average occupancy $\langle n \rangle$ and the standard deviation in occupancy $\sqrt{\langle n^2 \rangle - \langle n \rangle^2}$ as a function of the chemical potential $\mu$. \label{fg:fig2}}
\end{figure}

Is the grand canonical prescription accurate when the size of the solvation shell is comparable to the size of the particle ($R \sim r_p$)?  In Fig.~\ref{fg:fig2}, we show the distribution $p(n|R)$ of observing $n$ hard sphere solvent particles in a `solvation shell' of radius $R=2.2\times r_{p}$ around a distinguished hard sphere solute particle of the same size (black circles). The reduced density of the solvent is $8\rho r_p^3 = 0.9$. In order to ensure efficient sampling of rarely occupied states, we employ the expanded ensemble technique developed by Merchant et al.~\citep{merchant2011water}. The average number of solvent particles in the solvation shell is $\langle n \rangle \approx 4.5$.

Next, in order to find the best grand-canonical description of the solvation shell, we conducted grand canonical Monte Carlo simulations in a solvation shell of radius $R = 2.2\times r_{\rm p}$ for $\mu \in [-10, 5]$ with an interval of $\delta \mu = 0.05$ (see inset for $\langle n \rangle$ and $\sqrt{\langle n^2 \rangle - \langle n \rangle^2}$ as a function of $\mu$). Then, we chose the value of $\mu^{*} \approx -0.6$ that reproduced the average number of particles $\langle n \rangle \approx 4.5$ in the solvation shell.  The dashed blue line shows the grand canonical estimate of the number distribution $p_{\mu = \mu^{*}}(n|R)$.  Notably,  the grand canonical ensemble predicts a distribution with a  lower variance $\langle n^2 \rangle - \langle n \rangle^2$ in the occupancy statistics compared to the explicit simulation of hard sphere particles. 

What factors lead to these enhanced number fluctuations in the solvation shell compared to the grand-canonical ensemble? For molecular-sized solvation shells, local density fluctuations in the solvent are of the same magnitude as the occupancy of the solvation shell itself. Moreover, the density fluctuations outside the solvation shell will depend on the occupancy of the solvation shell. This implies that the work required to transfer  a solvent particle across the boundary of the solvation shell will depend on both the local density of solvent particles just outside the solvation shell as well as the density of solvent particles in the solvation shell~\citep{merchant2009thermodynamically,asthagiri2010ion,bansal_thermodynamics_2017}. As a result, a single chemical potential cannot represent the exchange of solvent particles between the solvation shell (`system') and the rest of the solvent (`bath'). From the point of view of statistical inference, the grand canonical distribution (Eq.~\ref{eq:gc0})  specifies all moments of $n$ with a single parameter $\mu$. Consequently the mean $\langle n \rangle$ and the variance $\langle n^2 \rangle - \langle n \rangle^2$ are coupled to each other through their dependence on $\mu$. In contrast, the mean and variance of $p(n)$ in Fig.~\ref{fg:fig2} can vary independent of each other.

How do we capture these enhanced number fluctuations in the solvation shell? Based on our previous work with the canonical ensemble~\citep{dixit2013maximum,dixit2015detecting}, we propose a superstatistical generalization. We hypothesize that a small system that exchanges particles with a surrounding medium can be represented by a system that is in contact with multiple baths, each characterized by a unique chemical potential $\mu$. Let $P(\mu)$ be the probability distribution over the baths. We can obtain the distribution over the number of solvent particles by marginalizing the variation over bath chemical potentials.
\begin{eqnarray}
	p_{\rm ss}(n|R) &=& \int P(\mu) \times p_\mu(n|R) d\mu \label{eq:2}
\end{eqnarray}
In Eq.~\ref{eq:2}, $p_\mu(n|R)$ is given by Eq.~\ref{eq:gc0}.  

What is the functional form of $P(\mu)$? One numerical approach, inspired by research in image processing~\citep{gull1984maximum}, is to constrain the $L_2$ error between the observed distribution $p(n|R)$ and the predicted distribution $p_{\rm ss}(n|R)$ while maximizing the entropy of $P(\mu)$. While this numerical approach can lead to accurate predictions~\citep{gull1984maximum}, the numerically inferred distribution $P(\mu)$ offers little physical clarity. Another, more conceptual approach is to motivate the functional form of $P(\mu)$ using first principles. Previously, we have shown using maximum entropy arguments that in a superstatistical generalization of the canonical ensemble~\citep{dixit2013maximum}, the distribution of inverse temperatures $P(\beta)$ can be described as an inverse gamma distribution.

Unfortunately, this functional form is not suitable for $P(\mu)$. This is because while inverse temperature $\beta$ for classical systems is always positive, chemical potential can take both positive and negative values. Notably, the inverse gamma distribution does not support negative arguments. However, the activity $z = \exp (\mu)$ only takes positive values. In this work, as a first guess,  we assume that bath chemical activities $z = \exp (\mu )$  are distributed as an inverse gamma distribution.  Thus, we assume that the bath chemical potential $\mu$ is distributed as 
\begin{eqnarray}
P(\mu) = \frac{  e^{\left (  -\lambda_1e^{-\mu}  - \lambda_2\mu \right )}}{\Gamma(\lambda_2)\lambda_1^{-\lambda_2}}. \label{eq:pmu}
\end{eqnarray} 
As we see below, this particular functional form accurately describes the solvent number fluctuations around a solute molecule. In the future, we would like to explore the relationship between $P(\mu)$ and the nature of system-bath interactions. 

\begin{figure}
	\includegraphics[scale=0.5]{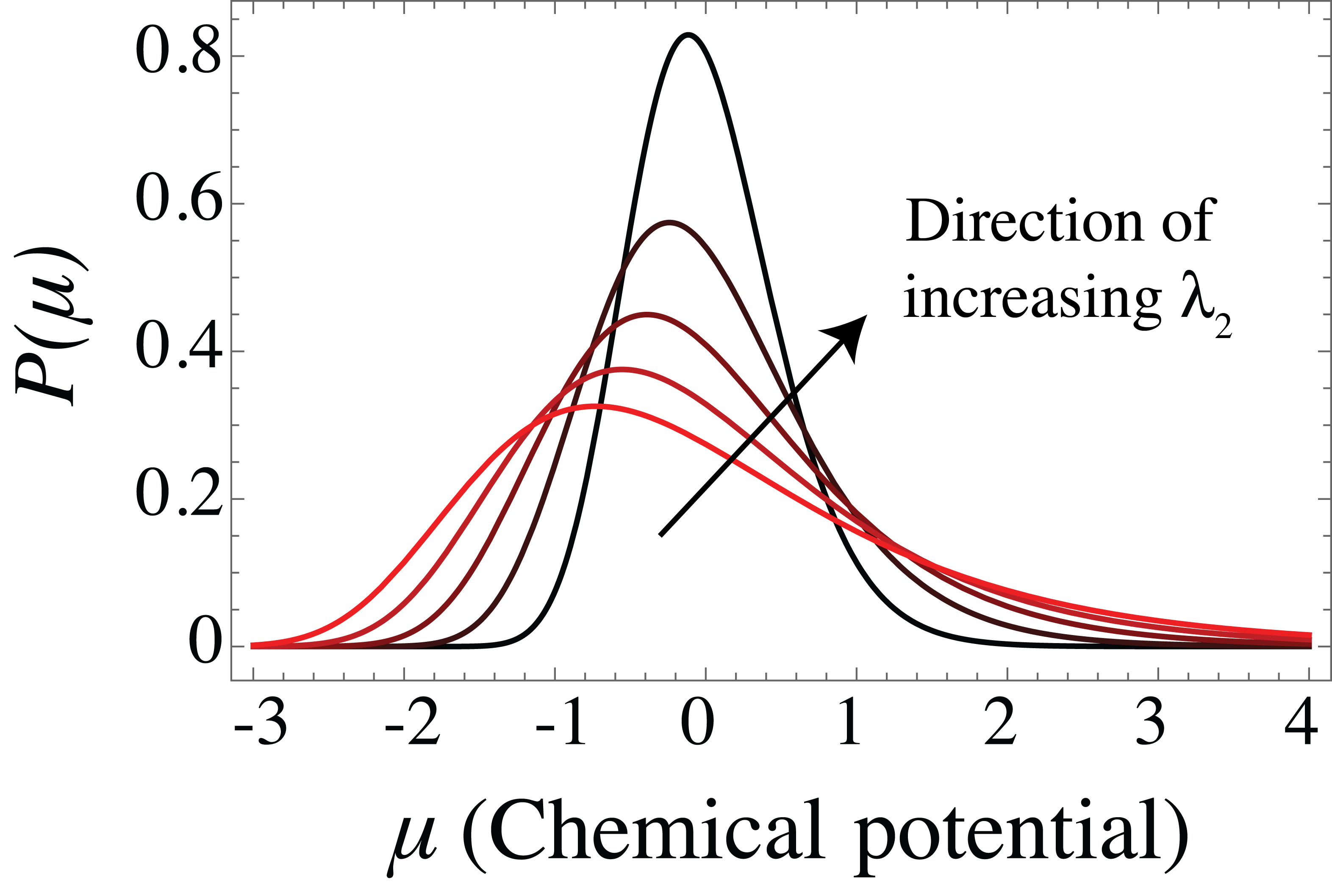}
	\caption{The distribution $P(\mu)$ as described by Eq.~\ref{eq:pmu} for different values of $\lambda_1$ and $\lambda_2$.  All shown distributions have the same mean $\langle \mu \rangle = 0$ and increasing standard deviation from black ($\sigma = 0.5$) to red ($\sigma = 1.5$) in steps of $\delta \sigma = 0.25$. The values of $\lambda_1$ and $\lambda_2$ for each of the distributions are calculated numerically by solving for the mean and the standard deviation. \label{fg:fig3}}
\end{figure}

Before we investigate whether $P(\mu)$ in Eq.~\ref{eq:pmu} can capture the solvent number fluctuations around the solute molecule, let us inspect its behavior. In Fig.~\ref{fg:fig3}, we plot different cases of $P(\mu)$. All shown distributions are constrained to have the same mean ($\langle \mu \rangle = 0$) and increasing standard deviation from black ($\sigma = 0.5$) to red ($\sigma = 1.5$) in steps of $\delta \sigma = 0.25$. 

\begin{figure*}
	\includegraphics[scale=0.65]{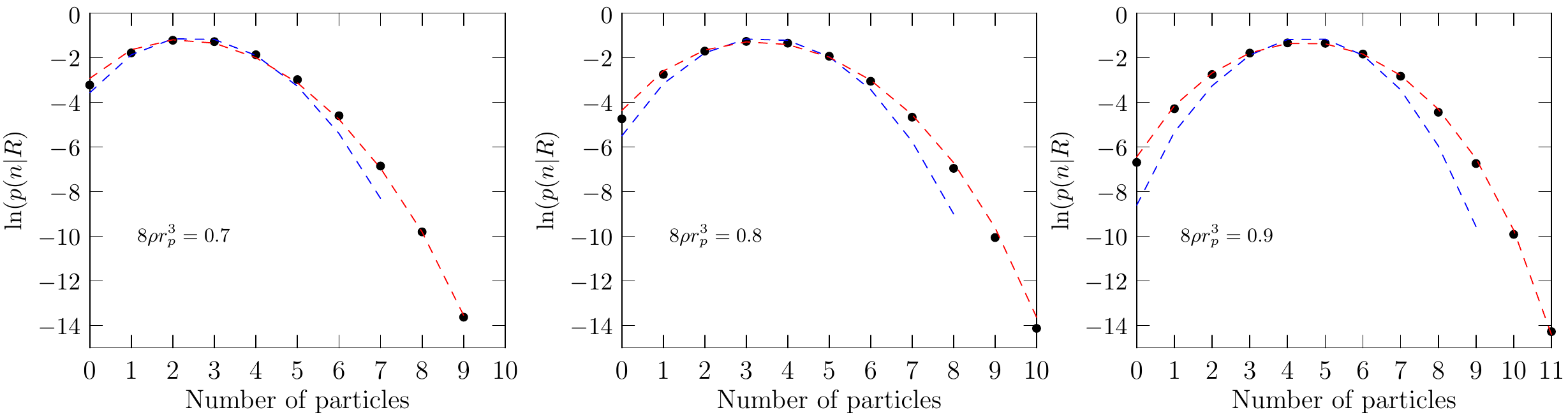}
	\caption{The probability $p(n|R)$ of observing $n$ solvent particles in a solvation shell of radius $R=2.2\times r_p$ of a solute particle of size $r_p$. We show $p(n|R)$ at reduced densities $8\rho r_p^3 = 0.7, 0.8$, and 0.9  respectively (black circles). The grand-canonical prediction $p_{\mu=\mu^{*}}(n|R)$ at each density is shown with dashed blue lines. The superstatistical prediction $p_{\rm ss}(n|R)$ is shown with dashed red lines.    \label{fg:fig4}}
\end{figure*}

Next, we test whether Eq.~\ref{eq:2} accurately capture the number statistics in the solvation shell. In Fig.~\ref{fg:fig4}, we show $p(n|R)$ at various reduced densities $8\rho r_p^3 = 0.7, 0.8, $ and 0.9. At each reduced density we find the chemical potential $\mu^{*}$ of the grand canonical ensemble that matches the mean occupancy $\langle n \rangle$. From Fig.~\ref{fg:fig4}, it is clear that for molecular-sized solvation shells, the occupancy statistics predicted by the grand canonical ensemble (Eq.~\ref{eq:gc0}) cannot capture the distribution of the number of solvent particles in the solvation shell.

Using the probabilities $p_\mu(n|R)$ of observing $n$ solvent particles in the solvation shell in grand canonical simulations, at each reduced density $8\rho r_p^3$, we numerically determined (using a simulated annealing scheme) the parameters $\lambda_1$ and $\lambda_2$ that lead to the lowest error when comparing $\log p(n|R)$  and $\log p_{\rm ss}(n|R)$. The dashed red lines show the predicted probability $p_{\rm ss}(n|R)$ using Eq.~\ref{eq:2}. Remarkably, the superstatistical distribution $p_{\rm ss}(n|R)$ can capture the entire distribution of solvent occupancy numbers very well for multiple solvent densities. In contrast, the grand canonical distribution of Eq.~\ref{eq:gc0} under-predicts the fluctuations in solvent occupancy numbers. Notably, a maximum entropy approach using a Gibbs prior or a flat prior on the occupancy distribution constrained by the mean occupancy and the variance in occupancy  also fails to capture the $p(n|R)$ distribution~\citep{hummer1996information,paliwal2006analysis}.

\begin{figure}
	\begin{center}
	\includegraphics[scale=0.5]{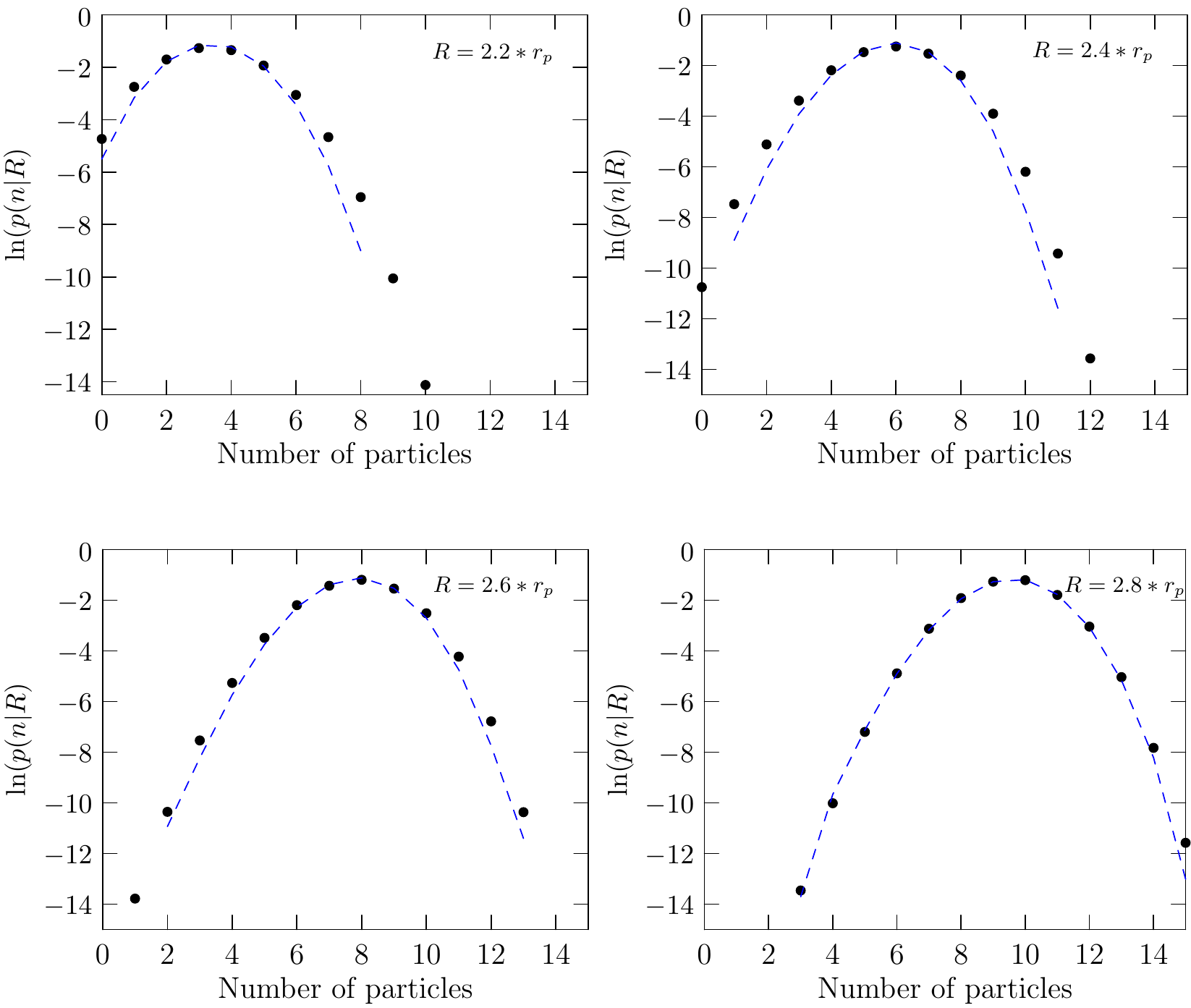}
	\end{center}
	\caption{The probability $p(n|R)$ of observing $n$ solvent particles in a solvation shell of radius $R=2.2\times r_p, 2.4\times r_p, 2.6\times r_p, $ and $2.8\times r_p$ (black dots) and the corresponding grand canonical monte carlo estimate $p_{\mu=\mu^{*}}(n|R)$ (dashed blue line). The reduced density of the solvent is $8\rho r_p^3 = 0.8$ in all 4 cases.      \label{fg:fig5}}
\end{figure}

We next investigated the approach to the eventual `macroscopic'  grand canonical description.  Using the expanded ensemble technique, we estimated the number distribution $p(n|R)$ for solvation shells of size $R = 2.2\times r_p, 2.4\times r_p, 2.6 \times r_p, $ and $2.8 \times r_p$.  The reduced density of the solvent was fixed at $8\rho r_p^3 = 0.8$.  Next, for each solvation shell radius $R$, we performed grand canonical Monte Carlo simulations and estimated the  $\mu^*$ that reproduced the mean occupancy $\langle n \rangle$. In Fig.~\ref{fg:fig5} we show $p(n|R)$  in black dots and the corresponding grand canonical estimate $p_{\mu=\mu^{*}}(n|R)$ in dashed blue lines. As expected, we observe that  the discrepancies between $p(n|R)$ and the grand canonical estimate $p_{\mu=\mu^{*}}(n|R)$ decrease as $R$ increases. 

{\bf Discussion:}  The statistics of occupancy of the solvation shell of a solute molecule is of central importance in understanding hydration phenomena~\citep{asthagiri2010ion,merchant2009thermodynamically,merchant2011ion} as well as in understanding preference of `binding sites' in proteins and nucleic acids towards small molecules~\citep{dixit2009ion,asthagiri2010ion,dixit2011role,dixit2011thermodynamics}. The grand canonical framework, if applicable at the microscopic scale, would be a versatile framework to understand solvation phenomena. This is because it would allow us to understand the occupancy statistics of a solvation shell as a combination of free energy $F(n)$ of formation of solute-solvent clusters that is independent of the bulk solvent and the effect of the bulk solvent represented by a single parameter; the chemical potential. Unfortunately, our recent work~\citep{bansal_structure_2016,bansal_thermodynamics_2017} and the current study shows that the grand canonical framework is inadequate in describing solvation shells. However,  our superstatistical approach allows us to interpret the effect of the bulk medium not as a single number (the chemical potential) but as a distribution.

How do we reconcile our development with previous work on understanding solvation phenomena? The non-constant nature of the bath chemical potential is well-recognized in the quasi-chemical approach~\citep{beck2006potential}. We describe connections with two previous works. First is the work of Merchant and Asthagiri that describes the {\it molecular aufbau principle} of the thermodynamic reorganization of the solvation shell upon addition of subsequent solvent molecules~\citep{merchant2009thermodynamically,merchant2011ion}. Second is the work by Bansal et al. that developes explicit solvent-related corrections to the grand canonical picture~\citep{bansal_structure_2016,bansal_thermodynamics_2017}. 

To be concrete, let $p^0(n|R)$ denote the probability of observing $n$ solvent particles in the solvation shell in the absence of the solute and $G(n)$ denote the free energy of assembling $n$ solute molecules in the solvation shell {\it in the presence of the solvent} (note that $G(n)$ is different from $F(n)$ in Eq.~\ref{eq:gc0}).Merchant and Asthagiri showed that~\citep{merchant2009thermodynamically,merchant2011ion}
\begin{eqnarray}
G - G(0) =  G(n) + \log p(n|R) - \log p^0(n|R) \label{eq:qc}
\end{eqnarray}
where $G$ is the excess free energy of introducing the solute particle in the solution. From Eq.~\ref{eq:qc}, we can calculate the chemical potential $G(n) - G(n-1)$ of the $n^{\rm th}$ solvent particle
\begin{eqnarray}
 G(n) -  G(n-1) = \log \frac{p(n-1|R)\times p^0(n|R)}{p(n|R) \times p^0(n-1|R)}   \label{eq:qc2}
\end{eqnarray}
From Eq.~\ref{eq:qc2}, it is clear that the work required to insert a solvent particle in the solvation shell is {\it not} constant. The work depends on the current occupancy of the solvation shell. In contrast, for a macroscopic system, the work in Eq.~\ref{eq:qc2} will be independent of the occupancy of the solvation shell and equal to the chemical potential. Thus, on the one hand, the quasi-chemical approach allows a detailed calculation of the dependence of chemical potential on the occupancy. On the other hand, the presented work complementarily allows us to rationalize enhanced density fluctuations in the solvation shell in terms of fluctuations in the bath chemical potential.

An approach more closely related to the current work is that of Bansal et al.~\citep{bansal_structure_2016,bansal_thermodynamics_2017}. In that work, we analyzed the solvation shell for hard sphere solutes explicitly in terms of the interactions within the cluster around the solute, interactions of the cluster with the rest of the medium and the interactions for the bulk solvent molecules which are not in the cluster. Based on the partition function in the canonical ensemble for the system which has $n$ molecules within the spherical shell around the  solute, we obtained the probabilities $p(n|R)$ as (in the notation of the current work)
\begin{equation}
p(n|R)\sim p_{\mu_p}(n|R)\times e^{\Omega \sigma_n}\label{eq:pn_bansal}
\end{equation}
where $\mu_{p}$ is the excess chemical potential of the solvent particle  and $p_{\mu_p}(n|R)$ is given by Eq.~\ref{eq:gc0} with the $\mu$ equal to the chemical potential of the solvent. $\Omega \sigma_n$ approximately represents the field imposed by the bulk solvent on the solute-solvent cluster. The field was explicitly recognized as a surface interaction term that depends on the occupancy of the surface sites in the cluster around the solute and represented as follows
\begin{equation}
 \Omega \sigma_n = \zeta_1 \cdot  n^2+\zeta_2\cdot n. 
\end{equation}
In other words, Bansal et al.~\citep{bansal_structure_2016,bansal_thermodynamics_2017}  apply a correction to the grand canonical ensemble explicitly for each coordination state (see also Reiss and Merry~\citep{reiss1981upper}). Interestingly, Eq.~\ref{eq:pn_bansal} can also be derived using maximum entropy arguments with the grand canonical distribution $p_{\mu_p}(n|R)$ as the prior (with the $\mu$ equal to the chemical potential of the solvent) with mean occupancy and second moment of occupancy as constraints. Note that absence of the correction term in Bansal et al. is equivalent to having $P(\mu)$ as a Dirac Delta function centered around the solvent chemical potential in the current work.

In summary, the grand canonical ensemble is inadequate in describing fluctuations in number statistics of molecular-sized solvation shells. In this work, we modeled the correction to the grand canonical ensemble in the form of a super-statistical ensemble where the solvation shell is coupled with multiple solvent baths, each with its own chemical potential. We found that the superstatistical description accurately captures the statistics $p(n|R)$ of the number $n$ of solvent molecules in the solvation shell of radius $R$. As the size of the solvation shell increases compared to the size of the solute particle, the agreement with the grand canonical description is restored. In this work, we approximated the distribution $P(z)$ of the bath chemical activities  with an inverse gamma distribution (and correspondingly $P(\mu)$ by Eq.~\ref{eq:pmu}). However, relevant experimental constraints can inform the choice of $P(\mu)$ as well. For example, in case of the canonical ensemble, previously, we have used maximum entropy arguments to motivate the form of the distribution over bath temperatures~\citep{dixit2013maximum,dixit2015detecting}.

{\bf Acknowledgments}
AB, WGC, and DA gratefully acknowledge support from Robert A. Welch Foundation (Grant \# C-1241) to Professor Walter Chapman. WGC and DA additionally acknowledge support from the Abu Dhabi National Oil Company (ADNOC).

%

\end{document}